\documentclass[a4paper,11pt]{article}
\pdfoutput=1 % if your are submitting a pdflatex (i.e. if you have
\usepackage{jheppub} % for details on the use of the package, please
\usepackage[T1]{fontenc} % if needed

\usepackage{pdfsync}
\usepackage{mathptmx}
\usepackage[utf8]{inputenc}
\usepackage[T1]{fontenc}
\usepackage{slashed}
\usepackage{times}
\usepackage{amssymb,amsfonts,amsmath,amsthm}
\usepackage{dsfont,bbm}
\usepackage{dcolumn}
\usepackage{epsf}
\usepackage{graphicx}
\usepackage[caption=false]{subfig}
\usepackage{dsfont}
\usepackage{bm}
\usepackage{eucal}
\usepackage{slashed}
\usepackage[active]{srcltx}
\usepackage[usenames]{color}
\title{\boldmath Conformal Symmetry and Effective Potential: II. Evolution}

\author[a]{I.~V.~Anikin}

\affiliation[a]{Bogoliubov Laboratory of Theoretical Physics JINR, 141980 Dubna, Russia}

\emailAdd{anikin@theor.jinr.ru}

\abstract{We present the second part of the paper series devoted to the study of the multi-loop effective potential
evolution (or anomalous dimension)
in $\varphi^4$-theory using the  conformal symmetry.
In this paper,
we demonstrate that the conformal symmetry can still be useful for
the effective potential approach even in the presence of  the mass parameter.
To this goal, it is necessary to introduce the special treatment of the mass terms as a sort
of interaction in an asymptotical expansion of the generating functional.
The introduced vacuum $V_{z,x}$-operation is the main tool to the algebraic scheme
of anomalous dimension calculations.
It is shown that the vacuum $V_{z,x}$-operation
transforms the given Green functions to the corresponding vacuum integrations which generate the effective potential.}

\begin{document}
\maketitle
\flushbottom

\section{Introduction}

In the field-theoretical models with spontaneous symmetry breaking at the classical level,
the geometrical analysis of the Goldstone theorem, based on the interpretation of a given
potential minimum as a vacuum state (see for instance \cite{Peskin:1995ev}), plays an important
role for the different theoretical approaches. As well-known, the quantum corrections which have been taken into account
for the theoretical studies distort the geometrical picture. However, the use of the effective potential methods allows to
return to the classical geometrical analysis of the models with spontaneous symmetry breaking.

The effective potential is given by the corresponding vacuum diagram sets
which are obtained 
as a result of the stationary phase method applied to the generating functional.
In the most interesting cases, we need to include the massive parameter in the models.
As well-known, if the loop integrations are made from the massive (scalar) propagators,
the conformal symmetry is useless in principle.
However, within the effective potential approach, we are able to avoid the use of massive propagator appearing in the vacuum diagrams.
To this aim, we propose to treat the mass terms in Lagrangian together with the coupling constant term
as a special kind of interaction which is forming by the corresponding effective vertices in an asymptotical expansion
of the generating functional \cite{Vasilev:2004yr}.
Hence, the scalar propagators in the vacuum diagrams describing interactions become massless ones
and open a window for the conformal symmetry use.

Recently, it has been shown in a series of papers \cite{Braun:2018mxm, Braun:2020yib, Braun:2016qlg, Braun:2014owa, Braun:2013tva} that
the recurrent relations inspired by the use of conformal symmetry
simplify substantially the multi-loop derivation of anomalous dimensions (evolution equations).
We adhere the method proposed and developed by Braun and Manashov in \cite{Braun:2013tva}.
In the Braun-Manashov (BM) approach, the corresponding anomalous dimensions at the given
$\ell$-loop accuracy can be derived practically without direct calculations but using the algebraic recurrent
relations originated from the conformal symmetry properties.
In the paper, due to the vacuum $V_{z,x}$ operation introduced in \cite{Anikin:2023wkk},
it is demonstrated that the anomalous dimensions of effective potential
can be readily obtained from the anomalous dimension of the non-local operator
Green function.

The present paper should be considered as an exploratory one where the
main features of the approach have been schematically demonstrated.
Basically, we outline the main ideas which are necessary to calculate the effective potential in
the frame of
$\varphi^4$-theory with the mass parameters. 
We have also adopted the BM-approach for computations of the effective potential evolution (or the anomalous dimension)
for the most general $\ell$-loop case.
We stress that the critical regime has been used in the auxiliary model which is the 
massless $\varphi^4$ theory where the conformal symmetry gives a possibility
to simplify drastically the multi-loop calculations (this is Braun-Manashov's approach).
The results obtained in the auxiliary model are still useful for our case because,
in MS-like schemes, the anomalous dimensions do not really depend on the
space-time dimension. So, all expressions
derived in the $D$-dimensional (conformal) theory remain exactly the same for the theory in the standard 
$4$-dimensions. In other words, having considered the auxiliary theory at the critical point
we do not lose any information on the anomalous dimensions in the non-critical theory.

The presentation is organized as follows. Sec.~\ref{bsec:GenFun} presents the details of stationary phase method
to describe the generating functional and the effective potential as a sum of the vacuum diagrams with the different effective vertices.
In this section, the effective interaction vertices have been introduced as a result of the asymptotical expansion
applied to the corresponding functional integration.
The way is shown to single out the Braun-Manashov (BM) correlators from the massless loop integrations.
Also, with the help of the special differentiation procedure,
the transition to the massless effective potential has been described.
In Sec.~\ref{Sec-BM}, the vacuum $V_{z,x}$ operation has been demonstrated.
In Sec.~\ref{Sec:CS-CB}, the evolution equation for the effective potential has been discussed.
For the reader convenience, App.~\ref{App:EE} contains the main details of the use of
the evolution (RG-like) equations, while App.~\ref{App:Delta} is devoted to the explanation on the
treatment of $\delta(0)$-singularity/uncertainty.

\section{Generating functional and effective potential in $\{\varphi^4\}_D$}
\label{bsec:GenFun}

\subsection{The standard generating functional for massive $\{\varphi^4\}_D$}
\label{subsec:GenFunM}

For the pedagogical reason,
we begin with the generating functional in the scalar $\varphi^4$ theory which leads to the effective action and potential.
Since the effective action differs from the effective potential by the (infinite) space-time volume,
$V\times T\sim \delta^{(D)}(0)$, in what follows we neglect this difference unless it leads to misunderstanding.
In a theory with interactions, the generating functional has the following form
(modulo the normalization constants denoted as {\it n.c.}):
\begin{eqnarray}
\label{GenFun-1}
&&{\mathbb Z}[J]\stackrel{n.c.}{=}e^{iS_I(\frac{\delta}{\delta J})} {\mathbb Z}_0[J]=
\int ({\cal D}\varphi)\, e^{iS(\varphi)+i(J,\,\varphi)},
\\
\label{GenFun-1-2}
&&{\mathbb Z}_0[J]={\cal N}e^{(J,\Delta_F \, J)}=
\int ({\cal D}\varphi)\, e^{iS_0(\varphi)+i(J,\,\varphi)},
\end{eqnarray}
where $\Delta_F$ implies the Feynman propagator;
$S(\varphi)=S_0(\varphi)+S_I(\varphi)$ denotes the sum of free and interaction actions. For the sake of shortness,
we use the notation
\begin{eqnarray}
\label{Den-1}
(a,K b)=\int dz_1\, dz_2\, a(z_1)\, K(z_1,z_2)\,b(z_2).
\end{eqnarray}
In Eqns.~(\ref{GenFun-1}) and (\ref{GenFun-1-2}), the stationary phase method allows us to calculate
both ${\mathbb Z}_0[J]$ and ${\mathbb Z}[J]$.
For free and interactive theories, the stationary conditions are given by
\begin{eqnarray}
\label{st-1}
%&&
\frac{\delta}{\delta\varphi} \Big[ S_0(\varphi)+(J,\,\varphi)\Big]=0,
%\\
\quad
\frac{\delta}{\delta\varphi} \Big[ S(\varphi)+(J,\,\varphi)\Big]=0
\end{eqnarray}
with the solutions as $\varphi_0^{(0)}$ and $\varphi_0=\varphi_0^{(0)} + \lambda \varphi_0^{(1)} + O(\lambda^2)$, respectively.

Notice that for the effective potential we need the classical field $\varphi_c$ which is generally not equal to $\varphi_0$,
{\it i.e.} $\varphi_c=\varphi_0+\varphi_1$, but $\varphi_c-\varphi_0=O(\hbar)$.
The classical field obeys the equation
\begin{eqnarray}
\label{phi-c}
\big( \Box +m^2\big) \varphi_c(x) = J(x) -
\frac{1}{{\mathbb Z}[J]}\, \big[  \lim_{A\to 0} \frac{\partial}{\partial A}V_{\varphi} (-i\frac{\delta}{\delta\varphi} + A)\big]
\, {\mathbb Z}[J],
\end{eqnarray}
where
\begin{eqnarray}
\label{V-def}
&&V(\varphi)=\lambda \varphi^4/4!, \quad S_I=\int dx V(\varphi)=\int dx {\cal L}_I(x),
\nonumber\\
&& \big( \Box +m^2\big) \varphi_0^{(0)}(x) = J(x)
\end{eqnarray}
and
\begin{eqnarray}
\label{W-def}
&&\varphi_c(x) = \frac{\delta {\mathbb W}[J]}{\delta J}=\langle \hat\varphi \rangle^J, \quad {\mathbb Z}[J]=e^{i\, {\mathbb W}[J]},
\nonumber\\
&&\lim\limits_{J\to 0} \varphi_c(x) = \varphi_c=const.
\end{eqnarray}
Moreover, the connected generalizing functional ${\mathbb W}[J]$ is related to the effective action $\varGamma[\varphi]$ as
(the Legendre transformations)
\begin{eqnarray}
\label{W-G-def}
\varGamma[\varphi] = {\mathbb W}[J] - i(J,\,\varphi).
\end{eqnarray}

Generally speaking, we may attempt to calculate $\varphi_1$ directly, however it is not necessary. Indeed,
we can assume that the current $J_1$ fulfils the equation
\begin{eqnarray}
\label{St-j1}
\frac{\delta S(\varphi)}{\delta\varphi}\Big|_{\varphi_c} +J_1=0, \quad \delta J=J - J_1.
\end{eqnarray}
With the above-defined current $J_1$, the stationary phase method applied to ${\mathbb Z}[J]$ gives the following
expression (here $\eta=\varphi - \varphi_c$)
\begin{eqnarray}
\label{Z-st-ph}
{\mathbb Z}[J]&=& e^{iS(\varphi_c) + i (J,\,\varphi_c)} \int ({\cal D}\eta)\,\,
\text{exp}\Big\{
\frac{i}{2!}\int dz_1 dz_2 \frac{\delta^2 S(\varphi)}{\delta\varphi(z_1)\delta\varphi(z_2)}\Big|_{\varphi_c} \eta(z_1) \eta(z_2)
\nonumber\\
&&
+\frac{i}{3!}
\int dz_1 .. dz_3 \frac{\delta^3 S(\varphi)}{\delta\varphi(z_1)..\delta\varphi(z_3)}\Big|_{\varphi_c} \eta(z_1).. \eta(z_3) + ...
\Big\}.
\end{eqnarray}

%%%%%%%%%%%%%%%%%%%%%%%%%%%%%%%%%%
\subsection{The masslessness procedure}
\label{SubSec-0}
%%%%%%%%%%%%%%%%%%%%%%%%%%%%%%%%%%%%%%%%%%%%%%%%%%%%%%%%

Introducing
$m^2_e(\varphi_c)=m^2+\lambda\varphi_c^2/2$ and assuming $\varphi_c=const$,
after an expansion of the corresponding exponentials
we get the following series
\begin{eqnarray}
\label{Z-st-ph}
&&{\mathbb Z}[J]= e^{iS(\varphi_c) + i (J,\,\varphi_c)} \int ({\cal D}\eta)
 e^{-\frac{i}{2}(\eta,\, \Box\eta)} \,\,\Big\{1 -
\frac{i}{2!}m^2_e(\varphi_c) \int dz \eta^2(z) + ... \Big\}
\nonumber\\
&&
\times
\Big\{1 -
\frac{i}{3!}
\lambda\varphi_c \int dz \eta^3(z) + ... \Big\}
\, \Big\{1 -
\frac{i}{4!}
\lambda\int dz \eta^4(z) + ... \Big\}
\end{eqnarray}
which should actually be considered as an asymptotical series.
The expansion of the mass exponential in a series (which actually transforms 
the mass from the propagator into the vertex) is valid if we are focusing on the UV-regime of 
loop integrations giving the needed anomalous dimensions. 

Hence, we generate the Feynman rules where we deal with
three types of interactions represented by the following
vertices
\begin{eqnarray}
\label{ver-1}
   &(a)&\Rightarrow  m^2_e(\varphi_c) \eta^2
\stackrel{\text{def}}{=} \lambda^{(a)} \eta^2 ;
\nonumber\\
   &(b)& \Rightarrow \lambda\varphi_c \eta^3
\stackrel{\text{def}}{=} \lambda^{(b)} \eta^3;
\nonumber\\
   &(c)&\Rightarrow \lambda\eta^4,
\end{eqnarray}
and all inner lines correspond to the scalar {\it massless} propagators.
Notice that in Eqn.~(\ref{ver-1}) the vertices $(a)$ and $(b)$ should be treated
as effective ones, while $(c)$ is standard vertex in the theory under consideration.

It is important to stress that the presence of massless propagators in our consideration open a window
to use the advantages of conformal symmetry as we discuss below
(all necessary details regarding the different aspects of the usage of conformal symmetry can be found in \cite{Braun:2003rp}).

Next, based on the generating functional (see Eqn.~(\ref{Z-st-ph})) and the Legendre transform
(see Eqn.~(\ref{W-G-def})) we can readily derive the expression
for the effective potential. Symbolically, we have
\begin{eqnarray}
\label{Eff-Pon-G-1}
\varGamma[\varphi_c]=S(\varphi_c) + \ln\big[ (\text{det}\,\widehat{\Box})^{-1/2}\big]
+ \,
\Big\{  n\text{-loop connected diagrams} \Big\}.
\end{eqnarray}
Here, the second term corresponds to the $1$-loop standard diagram and does not contribute in the massless case.
The third term in Eqn.~(\ref{Eff-Pon-G-1}) involves the full set of the connected diagrams
which can be grouped as the following
\begin{itemize}
\item the standard diagrams in $\{\varphi^4\}_D$ with the $[\lambda]^n$-vertices only;
\item the non-standard diagrams of type-$I$ with the $[\lambda^{(a)}]^n$-vertices only;
\item the non-standard diagrams of type-$II$ with the $[\lambda^{(b)}]^{2n}$-vertices only;
\item the diagrams of type-$III$ with the mixed vertices as
$[\lambda^{(a)}]^{n_1} [\lambda^{(b)}]^{n_2} [\lambda]^{n_3}$.
\end{itemize}
Notice that the standard vacuum diagrams with $[\lambda]^n$-vertices do not depend on $\varphi_c$ and, therefore,
they can be omitted at the moment.

At $\ell$-loop accuracy, every of connected diagrams contributing to the effective action of Eq.~(\ref{Eff-Pon-G-1})
contains the singular and finite parts. As usual, the singular parts should be eliminated by the corresponding counterterms
within the certain renormalization procedure resulting in the appearance of dimensional parameter (scale) $\mu$.
We remind that $\mu$ is related to some subtraction point.
The evolution of effective action/potential with respect to the different scale choice is governed by the corresponding
anomalous dimension (or Hamiltonian which has a form of the corresponding integral operator).
That is, we ultimately deal with the following effective action, see App.~\ref{App:EE},
\begin{eqnarray}
\label{EP-EE-1}
\varGamma[\varphi_c] = \sum_n \varGamma_n[\varphi_c]\equiv
\sum_n a_n \varGamma_n(0) \varphi^n_c(x)=
\sum_{n=2,4} a_n \varGamma_n(0) \varphi^n_c(x)+ ....
\end{eqnarray}
and
\begin{eqnarray}
\label{EP-EE-2}
\varGamma_n[\varphi_c]\Big|_{\mu_1} =
\varGamma_n[\varphi_c] \Big|_{\mu_2} \, {\rm exp}\Big\{ \int_{\mu_2}^{\mu_1} (dt) \gamma_{\varGamma_{n}}\Big\}.
\end{eqnarray}
In Eqn.~(\ref{EP-EE-1}), $\varGamma_n(0)$ denotes the $1$PI (vertex) Green functions, $a_n$ implies the combinatory factors, see below.

%%%%%%%%%%%%%%%%%%%%%%%%%%%%%%%%%%
\subsection{The non-standard diagrams of type-$I$ }
\label{SubSec-I}
%%%%%%%%%%%%%%%%%%%%%%%%%%%%%%%%%%%%%%%%%%%%%%%%%%%%%%%%

\begin{figure}[t]
\centerline{\includegraphics[width=0.5\textwidth]{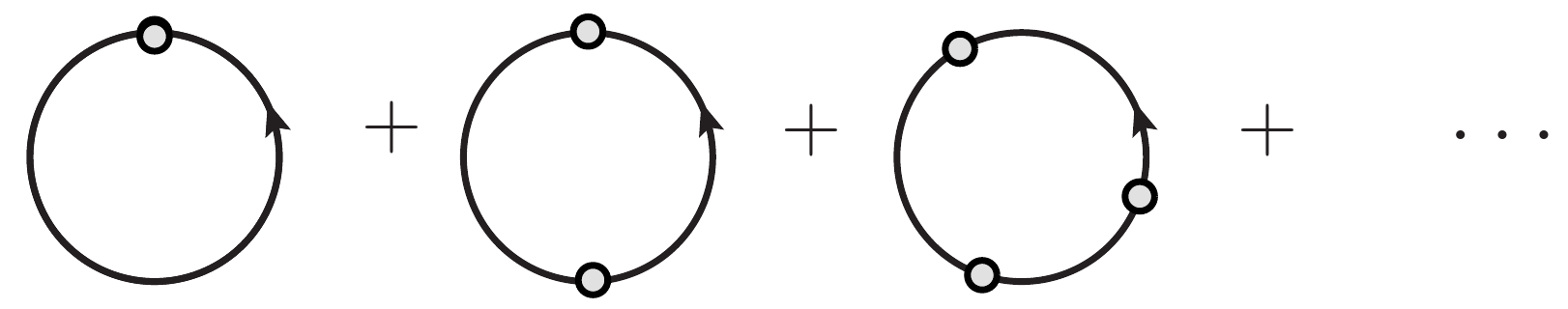}}
%\vspace{-0.5cm}
\caption{The diagrams of type-$I$ with only $\lambda^{(a)}$-vertices.}
\label{Fig-W-01}
\end{figure}

The non-standard diagrams of type-$I$, see Fig.~\ref{Fig-W-01}, contribute only to the one-loop approximation. For this type of diagrams, we have the
following representation ($D=4-2\varepsilon$)
\begin{eqnarray}
\label{diaI-1}
\varGamma^{(I)}[\varphi_c]=\sum_{n=1}^{\infty} \int (d^D k) \frac{[\lambda^{(a)}]^n}{(k^2)^n} =
\sum_{n=1}^{\infty} \frac{[\lambda^{(a)}]^n}{\Gamma(n)} \delta\big(n-D/2 \big)=
\frac{ [\lambda^{(a)}]^{2-\varepsilon}}{\Gamma(2-\varepsilon)} \delta(0).
\end{eqnarray}
Throughout the paper, the delta-function has been considered within the
sequential approach \cite{Antosik:1973} where the singularity/uncertainty of $\delta(0)$ should be treated
as the singularity of corresponding meromorphic function, {\it i.e.} $\lim_{\varepsilon\to 0}[ 1/\varepsilon]$,
see App.~{\ref{App:Delta}}
\cite{Anikin:2020dlh, Anikin:2023wkk}.
%
%
%%%%%%%%%%%%%%%%%%%%%%%%%%%%% FIGURE %%%%%%%%%%%%%%%%%%%%%%%%%%%%%%%%
%%%%%%%%%%%%%%%%%%%%%%%%%%%%%%%%%%%%%%%%%%%%%%%%%%%%%%%%%%%%%%%%%%%%%%%

%%%%%%%%%%%%%%%%%%%%%%%%%%%%%%%%%%
\subsection{The non-standard diagrams of type-$II$ }
\label{SubSec-II}
%%%%%%%%%%%%%%%%%%%%%%%%%%%%%%%%%%%%%%%%%%%%%%%%%%%%%%%%

\begin{figure}[t]
\centerline{\includegraphics[width=0.5\textwidth]{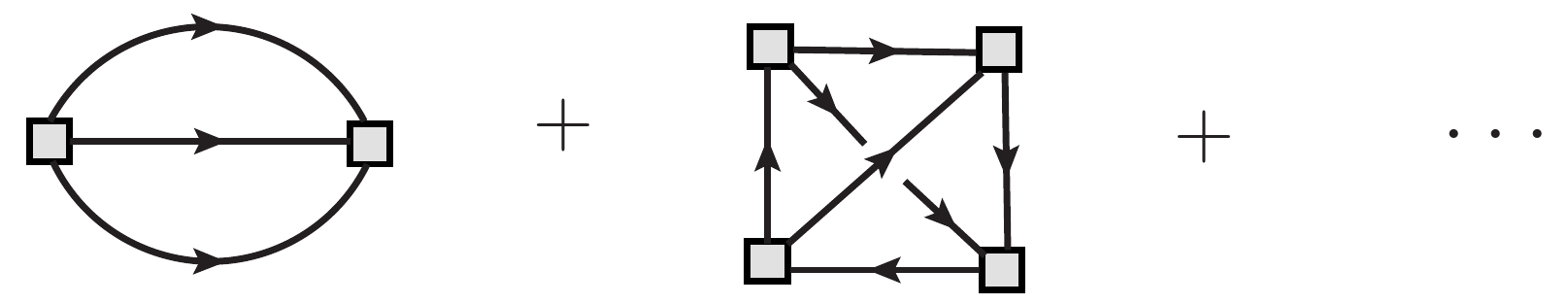}}
%\vspace{-0.5cm}
\caption{The diagrams of type-$II$ with only $\lambda^{(b)}$-vertices.}
\label{Fig-W-02}
\end{figure}

The non-standard diagrams of type-$II$ are given by the following set depicted in Fig.~\ref{Fig-W-02}.
One can see that among all these diagrams the non-zero contribution is stemmed from the three-loop box-like diagram that reads
\footnote{All $G$-functions are determined as in \cite{Grozin:2005yg}.}
\begin{eqnarray}
\label{diaII-1}
\varGamma^{(II)}[\varphi_c]=
G(1,1,1,1,1)\, \frac{ [\lambda^{(b)}]^{4}}{3\, \Gamma(2+ 2\varepsilon)} \delta(0),
\end{eqnarray}
where
\begin{eqnarray}
\label{G-4-1}
G(1,1,1,1,1)&=&\frac{2}{D-4} \Big\{ -(D-3) G^2(1,1)
\nonumber\\
&+&
\frac{(3D-8)(3D-10)}{D-4} G(1,1) G(1, 2-D/2)\Big\},
\\
\label{G-1-1}
G(n_1, n_2)&=&\frac{\Gamma(n_1+n_2-D/2)}{\Gamma(n_1)\Gamma(n_2)}
\frac{\Gamma(D/2-n_1)\Gamma(D/2-n_2)}{\Gamma(D-n_1-n_2)}.
\end{eqnarray}
Indeed, the general structure of the sum can be presented as
\begin{eqnarray}
\label{diaII-1-2}
\varGamma^{(II)}[\varphi_c]\sim \sum_{n=1}^{\infty}[\lambda^{(b)}]^{2n}
\, \delta\big(3 n - (n+1)D/2 \big).
\end{eqnarray}
Eqn.~(\ref{diaII-1-2}) shows that the only contribution originates from the case of $n=2$ that gives
$\delta(6-3D/2) \sim \delta(3 \varepsilon)$.

%
%
%%%%%%%%%%%%%%%%%%%%%%%%%%%%% FIGURE %%%%%%%%%%%%%%%%%%%%%%%%%%%%%%%%
%%%%%%%%%%%%%%%%%%%%%%%%%%%%%%%%%%%%%%%%%%%%%%%%%%%%%%%%%%%%%%%%%%%%%%%

%%%%%%%%%%%%%%%%%%%%%%%%%%%%%%%%%%
\subsection{The mixed diagrams of type-$III$ }
\label{SubSec-II}
%%%%%%%%%%%%%%%%%%%%%%%%%%%%%%%%%%%%%%%%%%%%%%%%%%%%%%%%

\begin{figure}[t]
\centerline{\includegraphics[width=0.5\textwidth]{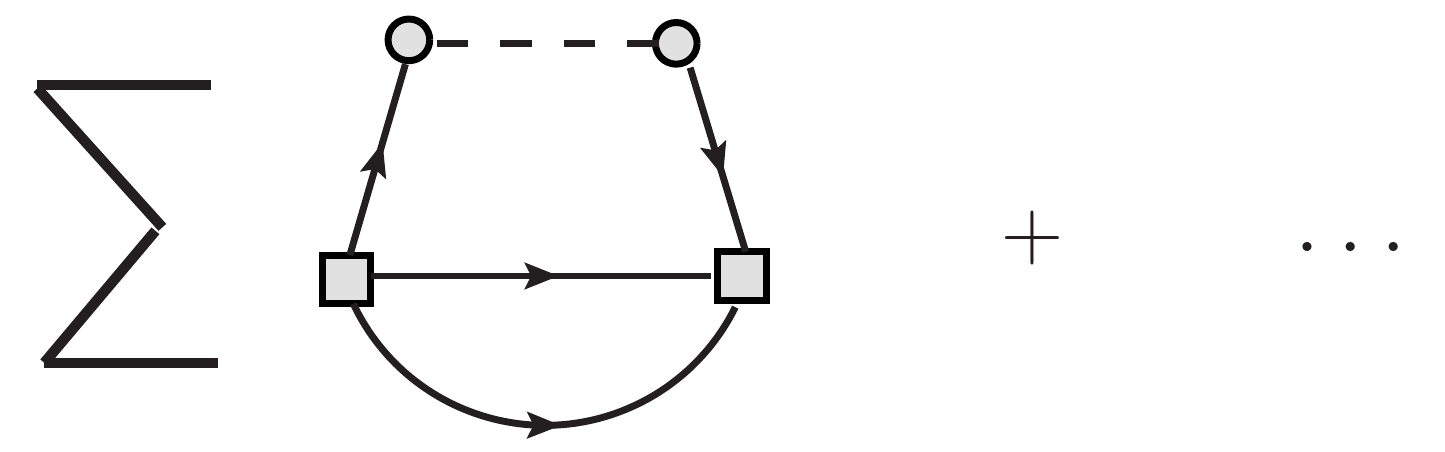}\quad \includegraphics[width=0.5\textwidth]{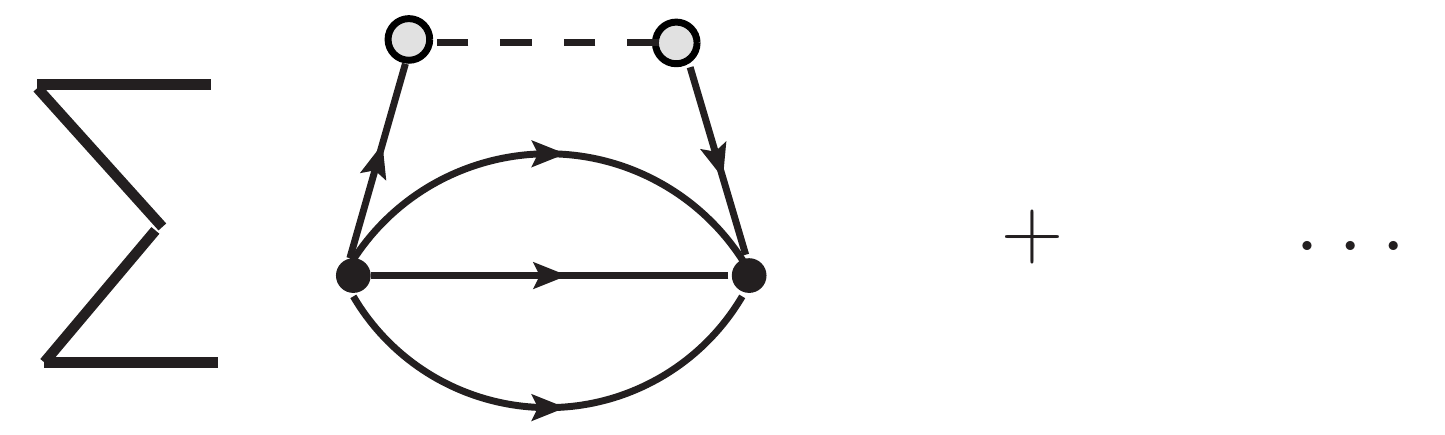}}
%\vspace{-0.5cm}
\caption{The diagrams of type-$III$: the left panel corresponds to the first class
with $\lambda^{(a)}$ and $\lambda^{(b)}$ vertices, the right -- the second class
with $\lambda^{(a)}$ and $\lambda$ vertices.}
\label{Fig-W-03}
\end{figure}

The mixed diagrams of type-$III$ can be aggregated into two classes. The first class of diagrams
with $n_1=n,\, n_2=2, \, n_3=0$, see the left panel of Fig.~\ref{Fig-W-03}, leads to
the two-loop contributions which are given by
\begin{eqnarray}
\label{diaIII-1}
\varGamma^{(III)}_{1}[\varphi_c]&=&[\lambda^{(b)}]^{2}\, \sum_{n=1}^{\infty} \int (d^D k) \frac{[\lambda^{(a)}]^n}{(k^2)^{n+1}}
\int  \frac{(d^D \ell)}{\ell^2 (\ell -k)^2}
\sim
[\lambda^{(b)}]^{2} \sum_{n=1}^{\infty}[\lambda^{(a)}]^n \delta\big(n+3 -D \big)
\nonumber\\
&=&
[\lambda^{(b)}]^{2}\, G(1,1) \frac{[\lambda^{(a)}]^{1-\varepsilon}}{\Gamma(2-\varepsilon)} \delta(0)
%\int (d^D k) \frac{[\lambda^{(a)}]}{(k^2)^{2}} \int \frac{(d^D \ell)}{\ell^2 (\ell -k)^2}.
\end{eqnarray}
The second class of diagrams with $n_1=n,\, n_2=0, \, n_3=2$, see the right panel of Fig.~\ref{Fig-W-03}, can be presented in the form of
three-loop integration as
\begin{eqnarray}
\label{Diam-1}
\varGamma^{(III)}_{2}[\varphi_c]&=&
[\lambda]^2 \,
\sum_{n=1}^{\infty} \int (d^D k)\frac{[\lambda^{(a)}]^n}{(k^2)^{n+1}} \, \int \frac{(d^D \ell)}{\ell^2}
\int \frac{(d^D p)}{p^2 (k+p-\ell)^2}
\nonumber\\
&\sim& [\lambda]^2 \, \sum_{n=1}^{\infty} [\lambda^{(a)}]^n \delta\big(n+4 -3D/2 \big)
\nonumber\\
&=& [\lambda]^2 \,G(1,1)\, G(1, \varepsilon)\,
\frac{[\lambda^{(a)}]^{2-3\varepsilon}}{\Gamma(2-\varepsilon)}\, \delta(0).
\end{eqnarray}
%
%
%%%%%%%%%%%%%%%%%%%%%%%%%%%%% FIGURE %%%%%%%%%%%%%%%%%%%%%%%%%%%%%%%%
%%%%%%%%%%%%%%%%%%%%%%%%%%%%%%%%%%%%%%%%%%%%%%%%%%%%%%%%%%%%%%%%%%%%%%%

For the sake of simplicity, we are only restricted by the order of $[\lambda]^4$ for
the connected diagrams. Though, the main features of calculations have a rather general character.

The contributions of $\varGamma^{(I)}[\varphi_c]$, $\varGamma^{(II)}[\varphi_c]$ and $\varGamma^{(III)}_1[\varphi_c]$
are uniquely fixed. That is, they only contribute to the definite order of $[\lambda]^k$ ($k=2,4,3$ respectively).
In contrast to these contributions, $\varGamma^{(III)}_{2}[\varphi_c]$ can involve the higher order of $[\lambda]^{k}$ with $k\ge 4$.

As usual, the singular parts of the diagram contributions given by $\varGamma^{(i)}[\varphi_c]$
generate the corresponding $Z$-factor needed for the mass and charge renormalizations \cite{Anikin:2023wkk}.
The anomalous dimensions are determined through the coefficients $c_1(\lambda)$ at the $1/\varepsilon$-singularities.
In the simplest case of lowest loop accuracy, it is not difficult to calculate the anomalous dimensions immediately.
However, the  highest loop (multi-loop) accuracy demands a lot of work.
We have found that the contribution of diagram given by $\varGamma^{(III)}_2[\varphi_c]$ to
the anomalous dimension can practically be computed algebraically based on the known anomalous dimension of the corresponding
non-local operator Green function $G^{(2)}_{\cal O}$ computed within BM-approach \cite{Braun:2013tva}.
It can be implemented due to the vacuum  $V_{z,x}$-operation \cite{Anikin:2023wkk}.

%%%%%%%%%%%%%%%%%%
\subsection{Massless effective potential in $\{\varphi^4\}_D$}
\label{SubSec:GenFun-NoMasses}
%%%%%%%%%%%%%%%%%%%%%%%%%%%%%%%%%%%%%%%%

To conclude this section,  let us discuss a formal transformation of action/potential with masses
to the massless (conformal-invariant) object.
In the case of $J\not= 0$, we consider the effective action/potential
\footnote{We remind that the effective potential is a part of the effective action which does not involve the derivatives over fields.
Therefore, if $\varphi_c=const$, the effective action is equivalent to the effective potential modulo $V\times T \sim \delta^{(D)}(0)$.}
given by the one-particle-irreducible (1PI) Green functions as
\begin{eqnarray}
\label{PI-GF}
\varGamma[\varphi_c] = \sum\limits_n \int (dx)_n\, \varGamma_{n}(x_1,...,x_n)\, \varphi_c(x_1) ... \varphi_c(x_n),
\end{eqnarray}
where the 1PI Green functions in $x$-space are transforming to the corresponding Green functions in $p$-space with the
nullified external momenta giving the vacuum diagrams, {\it i.e.}
\begin{eqnarray}
\label{GF-x-p}
\varGamma_{n}(x_1,...,x_n) \stackrel{{\cal F}}{=} \varGamma_{n}(p_1,...,p_n)\Big|_{p_i=0}\equiv  \varGamma_{n}(0),
\end{eqnarray}
where $ \stackrel{{\cal F}}{=}$ denotes the Fourier transform.

As above-mentioned, the theory under our discussion contains masses (or massive parameters) that destroy the conformal symmetry even at the
classical level.
Since we adhere the approach with small mass and coupling constant, it is legitimated to include the massive parameters in the vertices
forming the effective interactions, see Eqn.~(\ref{ver-1}). As a result, the scalar propagators in diagrams describing interactions are
massless ones.
We emphasize that, if the certain Feynman rules are fixed,  our method of diagram depicting 
is not much different from the usual technique with the standard $\lambda$-interaction vertex
in $\{\varphi^4\}_D$ and with the massive propagators.

Further, we focus on the simplest vacuum diagram with one $\lambda^{(a)}$ vertex, see Eqn.~(\ref{ver-1}),
 and one massless scalar propagator
(this is the so-called tadpole-like contribution). If we now remove the dimensionful vertices by the corresponding differentiation,
we can get the conformal invariant object determined by the massless scalar propagator, {\it i.e.}
\begin{eqnarray}
\label{Diff-1}
\frac{d \varGamma^{(a)}_{1}(0)}{d\lambda^{(a)}} = \varGamma^{(\eta^2)}_1(0)\stackrel{{\cal F}}{=} \Delta_F(0).
\end{eqnarray}
The other illustrative example is provided by
the Green function $\varGamma^{(a)(b)}_3(0)$ which corresponds to the vacuum diagram
with one $\lambda^{(a)}$ and two $\lambda^{(b)}$ vertices.
The loop integration of this diagram reminds the $2$-loop diagram in the massless $\{\varphi^4\}_D$ case.
If we again remove the dimensionful vertices in the similar way as in Eqn.~(\ref{Diff-1}), we obtain
\begin{eqnarray}
\label{Diff-2}
\frac{d^3  \varGamma^{(a)(b)}_3(0)}{d\lambda^{(a)}d\lambda^{(b)\,2}} =
\varGamma^{(\eta^2)(\eta^3)}_3(0),
\end{eqnarray}
where $\varGamma^{(\eta^2)(\eta^3)}_3(0)$ is the conformal invariant object as well.

%%%%%%%%%%%%%%%%%%%%%%%%%%%%%%%%%%
\section{From the Green functions to vacuum integrations with $V_{z,x}$-operator}
\label{Sec-BM}
%%%%%%%%%%%%%%%%%%%%%%%%%%%%%%%%%%%%%%%%%%%%%%%%%%%%%%%%

As mentioned in the preceding section, all vacuum integrations can be performed by direct calculations
\cite{Anikin:2020dlh, Gorishnii:1984te}.
However, in the case of $n_1=2$ (or, in other words, if the delta-function, appearing after the vacuum integration,
separates out the only term with $n_1=2$ in the full sum),
the method that is based on the manifestation of conformal symmetry can be applicable.

As an auxiliary theory, we use the massless $\varphi^4$ theory in the critical regime.
In \cite{Braun:2013tva}, it has been demonstrated that the certain constraints which are stemmed from
the conformal symmetry can be expressed in terms of the deformed generators of the collinear $SL(2)$ subgroup.
Notice that the conformal symmetry is manifested in the critical regime where $\beta(\lambda_*)=0$.
In particular, two generators $S_{-}$ and $S_{0}$ can be defined at all loops with the help of the
evolution kernel, while the special conformal
generator $S_{+}$ involves the nontrivial corrections
and it can be calculated order by order in perturbation
theory. Provided the generator $S_+$ is known at the
order of $(\ell -1)$ loop, the corresponding evolution kernel in
the physical dimension can be fixed to the $\ell$-loop accuracy (up to the terms
which are invariant regarding the tree-level generators).
In other words, BM-approach allows us to derive the corresponding anomalous dimensions at the given
$\ell$-loop accuracy practically without direct calculations but using the algebraic recurrent
relations originated from the conformal symmetry properties.
Since, in MS-scheme (or in the similar schemes), the anomalous dimensions are independent on the
dimension, the anomalous dimensions
derived in the $D$-dimensional (conformal) theory are exactly the same even for the theory in the standard
$4$-dimensions. So, the results derived in the auxiliary theory at the critical point
contain all information on the anomalous dimensions in the non-critical $4$-dimension theory.

In this section, we show that the advantages of BM-approach can be used
to derive algebraically the evolution of effective potential $\varGamma[\varphi_c]$ at any loop accuracy.
To this purpose, roughly speaking, we have to relate the effective potential to the corresponding Green functions
which have been studied in \cite{Braun:2013tva}.
It turned out that it becomes possible if
we introduce the vacuum $V_{z,x}$-procedure which transforms the usual Green functions to
the vacuum integrations.
Working with Eqn.~(\ref{Diam-1}), we define the $V_{z,x}$-procedure as  \cite{Anikin:2023wkk}
\begin{eqnarray}
\label{Diam-1-2}
\varGamma^{(n)}[\varphi_c]= \frac{1}{C^{(n)}(D)}V_{z,x} \Big\{
G^{(n)}_{\cal{O}}(x_1, x_2 ; z_1, z_2)\Big\},
\end{eqnarray}
where $C^{(n)}(D)$ denotes the combination of $\Gamma$-functions
\footnote{The exact representation of
this normalization can be found in \cite{Anikin:2023wkk}.}, and
\begin{eqnarray}
\label{Diam-1-2-3}
&&V_{z,x} \Big\{
G^{(n)}_{\cal{O}}(x_1, x_2 ; z_1, z_2)\Big\} \stackrel{\text{def}}{=}
\\
&&
[\lambda^{(a)}]^{3D/2-4}\,
\int d^D z_1\, d^D z_2 \Delta_F(z_1-z_2) \Big[
\int d^D x_1 d^D x_2 \delta(x_1-x_2) \widehat{\Box}_{x_2} \,
\Big\{ G^{(n)}_{\cal{O}}(x_1, x_2 ; z_1, z_2)\Big\} \Big].
\nonumber
\end{eqnarray}
In the interaction representation, the non-local operator Green function
$G^{(n)}_{\cal{O}}(x_1, x_2 ; z_1, z_2)$ to $[\lambda]^n$-order
reads
\begin{eqnarray}
\label{GF-1}
G^{(n)}_{\cal{O}}(x_1, x_2 ; z_1, z_2) =
\langle 0| T \eta(x_1) \eta(x_2)\, {\cal O}(z_1, z_2)
\Big( [\lambda] \int d^D y \eta^4(y)\Big)^n |0\rangle,
\end{eqnarray}
with ${\cal O}(z_1, z_2)=\eta(z_1)\eta(z_2)$
and $[\lambda]=\lambda/4!$.

On the other hand, $G^{(n)}_{\cal{O}}(x_1, x_2 ; z_1, z_2)$ can be written as
\begin{eqnarray}
\label{CS-ob-1}
G^{(n)}_{\cal{O}}(x_1, x_2 ; z_1, z_2) =
\langle {\cal O}(z_1, z_2)\rangle^{(n)}
\Big |^{ \eta(z^{\alpha_1}_{12}) \to \Delta_F(x_1-z^{\alpha_1}_{12})}_{\eta(z^{\alpha_2}_{21}) \to \Delta_F(z^{\alpha_2}_{21}-x_2)},
\end{eqnarray}
where the correlator of non-local operator is defined as
\begin{eqnarray}
\label{O-op-me}
\langle {\cal O}(z_1, z_2)\rangle^{(n)}=
\langle 0| T {\cal O}(z_1,z_2) \Big( [\lambda] \int d^D y \eta^4(y)\Big)^n |0 \rangle.
\end{eqnarray}
We emphasize that $\langle {\cal O}(z_1, z_2)\rangle$ is now the BM-like object which we need for our consideration.

For the sake of simplicity, our consideration begins with the $[\lambda]^2$-order, {\it i.e.} $n=2$.
In this case, using Eqns.~(\ref{GF-1}) and (\ref{CS-ob-1}), we can rewrite Eqn.~(\ref{Diam-1-2}) as
\begin{eqnarray}
\label{Diam-1-3}
\varGamma^{(2)}[\varphi_c]=
V_{z,x}\Big\{
\langle {\cal O}(z_1, z_2)\rangle^{(2)}
\Big |^{ \eta(z^{\alpha_1}_{12}) \to \Delta_F(x_1-z^{\alpha_1}_{12})}_{\eta(z^{\alpha_2}_{21}) \to \Delta_F(z^{\alpha_2}_{21}-x_2)} \Big\}.
\end{eqnarray}
For the non-local correlator $ {\cal O}(z_1, z_2)$, one can calculate the anomalous dimension
using the Braun-Manashov method \cite{Braun:2013tva}.

%%%%%%%%%%%%%%%%%%%%%%%%%%%%%%%%%%%%%%%%%%%%%%%%%
\section{Evolution kernel for effective potential}
\label{Sec:CS-CB}
%%%%%%%%%%%%%%%%%%%%%%%%%%%%%%%%%%%%%%%%%%%%%%%%%

In this section, we are going over to the discussion of the evolution equation for the
effective potential.

\begin{figure}[t]
\centerline{\includegraphics[width=0.25\textwidth]{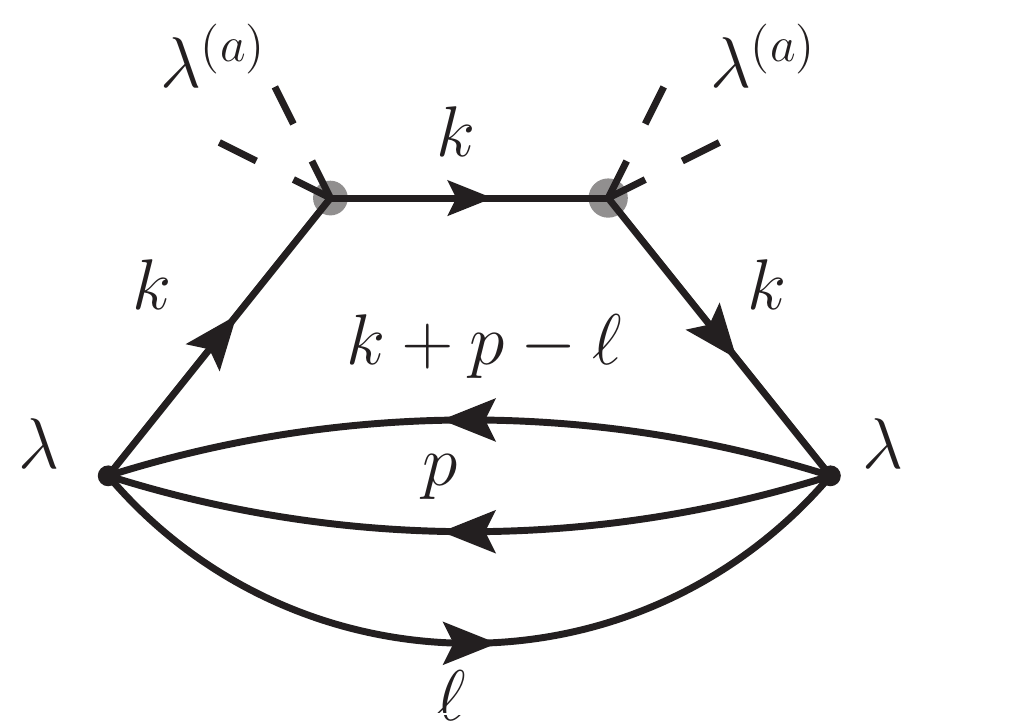}}
%\vspace{-0.5cm}
\caption{The diagram with $[\lambda]^2 [\lambda^{(a)}]^2$-vertices.}
\label{Fig-W-2}
\end{figure}

Let us now consider the diagram presented by the loop integration $\varGamma^{(2)}[\varphi_c]$ of Eqn.~(\ref{Diam-1-3}),
see Fig.~\ref{Fig-W-2}.
The conformal BM-object can be obtained by the differentiation as
\begin{eqnarray}
\label{Diam-1-2-3}
\overline{\varGamma^{(2)}[\varphi_c]}= \frac{\partial^2 \varGamma^{(2)}[\varphi_c]}{\partial \lambda^{(a)\, 2}}=
\overline{V}_{z,x}\Big\{
\langle {\cal O}(z_1, z_2)\rangle^{(2)}
\Big |^{ \eta(z^{\alpha_1}_{12}) \to \Delta_F(x_1-z^{\alpha_1}_{12})}_{\eta(z^{\alpha_2}_{21}) \to \Delta_F(z^{\alpha_2}_{21}-x_2)} \Big\}.
\end{eqnarray}
As mentioned, $ {\cal O}(z_1, z_2)$ of Eqn.~(\ref{Diam-1-3}) can be treated as a subject
of BM-approach. Schematically, our procedure of evolution kernel derivation can be described in the following way.
First, we calculate the anomalous dimension of $ {\cal O}(z_1, z_2)$ at the order of $[\lambda]^2$. We get
\begin{eqnarray}
\label{H-1}
\langle {\cal O}(z_1, z_2)\rangle^{(2)}  \Rightarrow \frac{1}{\varepsilon} \big[ \mathbb{H}^{(2)}_{12} {\cal O}\big] (z_1,z_2).
\end{eqnarray}
Then, we apply the $V_{z,x}$-operation to get the coefficient $c_2^\varGamma$ at $1/\varepsilon^2$-singularity in the
effective potential $\varGamma^{(2)}[\varphi_c]$, {\it i.e.} it reads
\begin{eqnarray}
\label{H-2}
c_1^{\cal O}=\big[ \mathbb{H}^{(2)}_{12} {\cal O}\big] \,\stackrel{V_{z,x}}{\Longrightarrow} \, c_2^{\varGamma}.
\end{eqnarray}
The last step is to use the corresponding pole relations written for the effective potential in order to obtain the
anomalous dimension (evolution kernel) for $\varGamma[\varphi_c]$, we have
\begin{eqnarray}
\label{H-3}
c_1^\varGamma = {\rm P}(c_2^\varGamma) \equiv \big[ \mathbb{H}^{(2)} \varGamma[\varphi_c] \big].
\end{eqnarray}
The operator ${\rm P}$ is entirely defined by the pole relations.
In its turn, the pole relations are stemmed from
the $\mu\partial_\mu$-differentiation of the effective potential $Z$-factors, $Z_m$ and $Z_\lambda$,
see \cite{Anikin:2023wkk}.

%
%
%%%%%%%%%%%%%%%%%%%%%%%%%%%%% FIGURE %%%%%%%%%%%%%%%%%%%%%%%%%%%%%%%%
\begin{figure}[t]
\centerline{\includegraphics[width=0.25\textwidth]{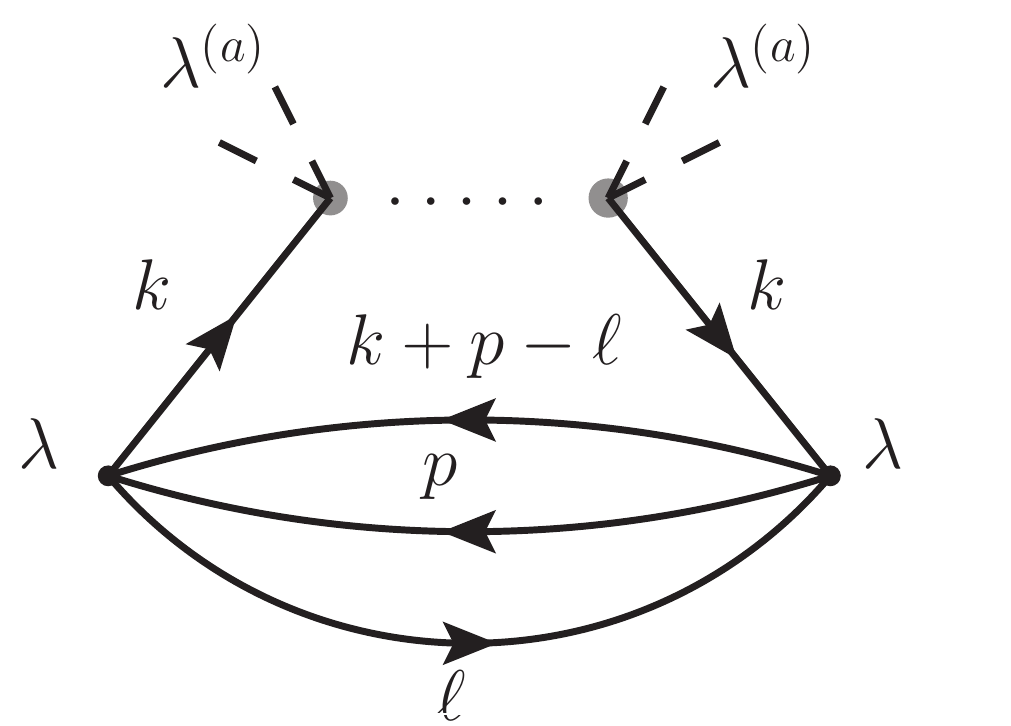}}
%\vspace{-0.5cm}
\caption{The diagram with $[\lambda]^2 [\lambda^{(a)}]^n$-vertices.}
\label{Fig-W-1}
\end{figure}
%%%%%%%%%%%%%%%%%%%%%%%%%%%%%%%%%%%%%%%%%%%%%%%%%%%%%%%%%%%%%%%%%%%%%%%

%
%
%%%%%%%%%%%%%%%%%%%%%%%%%%%%% FIGURE %%%%%%%%%%%%%%%%%%%%%%%%%%%%%%%%
%%%%%%%%%%%%%%%%%%%%%%%%%%%%%%%%%%%%%%%%%%%%%%%%%%%%%%%%%%%%%%%%%%%%%%%

%%%%%%%%%%%
\subsection{The Braun-Manashov method in a nutshell: the recurrent relations from the conformal symmetry}
\label{SubSec:CS-RR}
%%%%%%%%%%%%%%%%%%%%%%%%%%%

We are now in a position to discuss 
the main items of the method proposed and developed by Braun and Manashov \cite{Braun:2013tva}.
The method is based on
the recurrent relations inspired by the conformal symmetry.
In the paper, we merely use the finding of \cite{Braun:2013tva} and follow to their convention of notations.
Namely, since we work in the frame
at the critical point, {\it i.e.} $\lambda = \lambda_{*}$ and $\beta(\lambda_{*})=0$,
we can extend our symmetry to the dilatation and the space-time inversion which form
the well-known collinear $SL(2)$ subgroup of the conformal group.

As in \cite{Braun:2013tva}, we assume that the collinear conformal algebra can be realized by
the standard way with the help of operators  $\mathbb{L}_\pm$ and $\mathbb{L}_0$.
Obviously, a non-local operator can be considered as a generalizing function for a local operator.
For the renormalized operator we write that
\begin{eqnarray}
\label{Nonlon-loc}
[{\cal O}](z_1, z_2) =
\sum\limits_{Nk} \Psi_{Nk}(z_1,z_2)\, [{\cal O}]_{Nk}
\end{eqnarray}
where $\Psi_{Nk}(z_1,z_2)$ being homogeneous polynomials of degree $N + k$
meets the requirement $(z_i\partial_{z_i} + N-k)\Psi_{Nk}(z_1,z_2)=0$.
One can construct the adjoint representation of operators $\mathbb{L}_i$.
That is, instead of the generators $\mathbb{L}_i$ which act on the operator fields, see Eqn.~(\ref{non-loc-1}),
we can introduce the operators $S_\alpha$ which act on the coefficient functions $\Psi_{Nk}(z_1,z_2)$
\footnote{We address the reader to \cite{Belitsky:2005qn} to find a very detail and pedagogical description of this procedure.},
see Eqn.~(\ref{Nonlon-loc}).

The generators $S_\alpha$ also obey the standard commutation relations, we have
\begin{eqnarray}
[S_\pm, S_0] = \mp S_\pm, \quad  [S_+, S_-] = 2 S_0
\end{eqnarray}
with the following realization on the space of homogeneous polynomials
\begin{eqnarray}
\label{S-op}
&&
S_- \Psi_{Nk}(z_i) = - \Psi_{N k-1}(z_i), \,\, S_0 \Psi_{Nk}(z_i) = (j_N+k) \Psi_{N k}(z_i),
\nonumber\\
&&
S_+ \Psi_{Nk}(z_i) = (k+1)(2j_N+k) \Psi_{N k+1}(z_i)
\nonumber\\
&&\text{with } \quad S_\alpha = S^{(0)}_\alpha + \Delta S_\alpha.
\end{eqnarray}
In Eqn.~(\ref{S-op}), the operators $S_\alpha$ within a free theory have the forms of
\begin{eqnarray}
\label{S-form-1}
 S^{(0)}_- = - \sum\limits_i \partial_{z_i},
\quad
S^{(0)}_0 = \sum\limits_i z_i \partial_{z_i} + 2j,
\quad
S^{(0)}_+ =\sum\limits_i (z_i^2 \partial_{z_i} + 2j z_i),
\end{eqnarray}
while the interaction modifies the operators $S_\alpha$ by adding extra terms as
\begin{eqnarray}
\label{S-form-2}
\Delta S_- = 0,
\quad
\Delta S_0 = -\varepsilon +\frac{1}{2}\mathbb{H}(\lambda_{*}),
\quad
\Delta S_+ =\sum\limits_i z_i \Big( -\varepsilon + \frac{\lambda_{*}}{2}
\mathbb{H}^{(1)}\Big) + O(\varepsilon^2),
\end{eqnarray}
where $\mathbb{H}$ denotes the anomalous dimension (or, according to the terminology of \cite{Braun:2013tva}, Hamiltonian).

Following the Braun-Manashov approach, we remind that
 $[\mathbb{H}, S^{(0)}_\alpha] \not= 0$ beyond the
leading order in the interacting theory. However, as it is pointed out in Eqn.~(\ref{S-op}),  the generators $S_\alpha$
can be defined as a sum $S^{(0)}_\alpha+\Delta S_\alpha$ which satisfy the canonical $sl(2)$ commutation
relations for the theory at the critical coupling in non-integer dimensions.
It is worth to notice that the commutation relations impose certain self-consistency relations on
the corrections $\Delta S_\alpha$.

Further, having expanded the relation $[S_+,\mathbb{H}(\lambda_{*})] = 0$ in powers of $\lambda_{*}$,
the following relations take finally the form of (see \cite{Braun:2013tva})
\begin{eqnarray}
\label{reccur-1-w1}
&&[S^{(0)}_+ , \mathbb{H}^{(1)}] = 0 ,
\quad
[S^{(0)}_+ , \mathbb{H}^{(2)}] = [\mathbb{H}^{(1)}, \Delta S^{(1)}_+],
\nonumber\\
&&
[S^{(0)}_+ , \mathbb{H}^{(3)}] = [\mathbb{H}^{(1)}, \Delta S^{(2)}_+]
+ [\mathbb{H}^{(2)}, \Delta S^{(1)}_+]  \quad \text{etc.},
\end{eqnarray}
where
\begin{eqnarray}
\label{Exp-lambda}
\Delta S_+ =\sum\limits_{k=1}^\infty
\lambda^k_{*} \Delta S^{(k)}_+ , \quad
\mathbb{H} = \sum\limits_{k=1}^\infty \lambda^k \mathbb{H}^{(k)}.
\end{eqnarray}
The relations of Eqn.~(\ref{reccur-1-w1}) show that if the anomalous dimension $\mathbb{H}^{(k)}$
is known at the given $\ell$-loop accuracy together with the representation for the corresponding deformed operator $\Delta S^{(m)}_+$,
the anomalous dimension $\mathbb{H}^{(k+1)}$ at the given $(\ell +1)$-loop accuracy can be derived almost algebraically.
For our goal, it means that with the help of $V_{z,x}$-operation we can also readily derive the evolution kernel for the effective potential
$\varGamma[\varphi_c]$.

\subsection{Demonstration of the method}
\label{SubSec:Demo}

\begin{figure}[t]
\centerline{\includegraphics[width=0.3\textwidth]{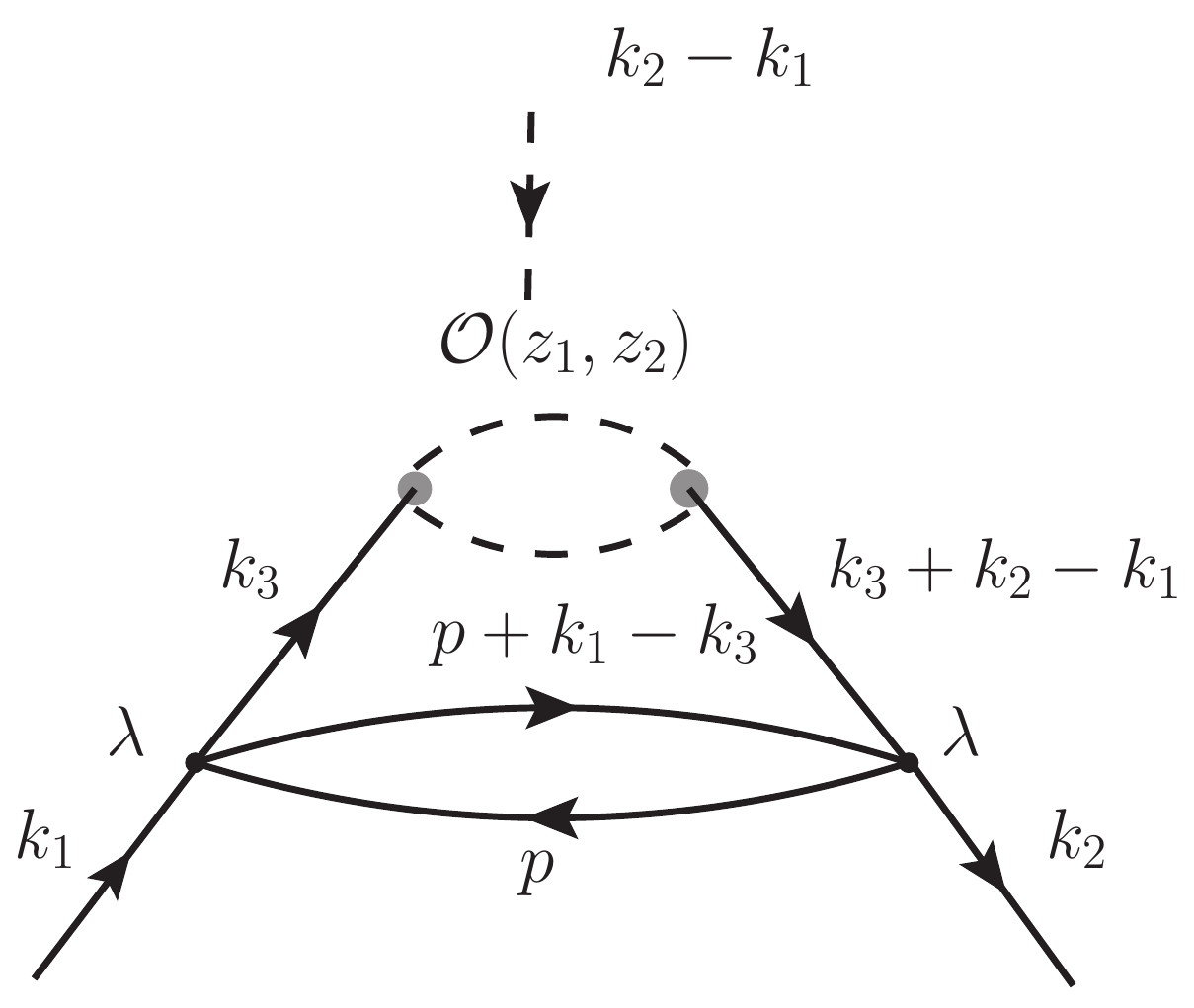}}
%\vspace{-0.5cm}
\caption{The diagram of $G^{(2)}_{\cal{O}}(x_1, x_2 ; z_1, z_2)$ at the order of $[\lambda]^2$.}
\label{Fig-W-3}
\end{figure}

Let us suppose that the one-loop evolution kernel $\mathbb{H}^{(1)}$ has been someway calculated
\footnote{We omit the details of calculations because they are of absolutely standard nature.}.
Using the mentioned recurrent relations, see Eqn.~(\ref{reccur-1-w1}),
after some algebra one can derive that
the evolution kernel for the two-loop accuracy takes a form of \cite{Braun:2013tva}
\begin{eqnarray}
\label{HK-2}
\mathbb{H}^{(2)}=  \mathcal{H}^+  + \mathbb{F}\big(\mathcal{H}^{(d)},  \mathcal{V}^{(d, 1)}\big)
%3  \mathcal{H}^{(d)} + \mathcal{H}^+ \Pi_0 + \mathcal{V}^{(d, 1)} - \frac{1}{6},
\end{eqnarray}
where $\mathbb{F}(...)$ implies the combinations which do not finally contribution to the vacuum integrations,
and
\begin{eqnarray}
\label{Int-H-3}
\mathcal{H}^+ \equiv \big[ \mathcal{H}^+ {\cal O}\big] (z_1,z_2) =
\int_{0}^{1} d\alpha_1 \int_{0}^{\overline{\alpha}_1} \frac{d\alpha_2}{1-\alpha_{12}} {\cal O}(z_{12}^{\alpha_1}, z_{21}^{\alpha_2}),
\end{eqnarray}
where $\alpha_{12...n}=\alpha_1+....+\alpha_n$.
This contribution corresponds to the diagram depicted in Fig.~\ref{Fig-W-3}.
Due to the relative simplicity,  it can be directly calculated without the usage of the recurrent relations.
However, in the case of higher loop corrections, the recurrent relations are very useful because they replace rather
complicated direct calculations by the almost algebraic calculations.

Next, we apply our $\overline{V}_{z,x}$-operation, see Eqn.~(\ref{Diam-1-2-3}),
in order to get the coefficient $c_2^{\overline{\varGamma}}$ for the effective potential.
Then, we get 
\begin{eqnarray}
\label{Vac-c2}
c_2^{\overline{\varGamma}}=
\overline{V}_{z,x}\Big\{
 \big[ \mathcal{H}^+ {\cal O}\big] (z_1,z_2)
\Big |^{ \eta(z^{\alpha_1}_{12}) \to \Delta_F(x_1-z^{\alpha_1}_{12})}_{\eta(z^{\alpha_2}_{21}) \to \Delta_F(z^{\alpha_2}_{21}-x_2)} \Big\}.
\end{eqnarray}
It now remains to insert this representation into Eqn.~(\ref{H-3}) to derive the needed anomalous dimension for the effective
potential.

%
%
%%%%%%%%%%%%%%%%%%%%%%%%%%%%% FIGURE %%%%%%%%%%%%%%%%%%%%%%%%%%%%%%%%
%%%%%%%%%%%%%%%%%%%%%%%%%%%%%%%%%%%%%%%%%%%%%%%%%%%%%%%%%%%%%%%%%%%%%%%

%%%%%%%%%%%%%%%%%%%%%%%%%%%%%%%%%%%%%%%%%%%%%%%%%
\section{Conclusion}
\label{Sec:Con}
%%%%%%%%%%%%%%%%%%%%%%%%%%%%%%%%%%%%%%%%%%%%%%%%%

We have outlined a new approach to calculate the multi-loop effective potential
evolutions in $\varphi^4$-theory using the  conformal symmetry.
In this paper,
we have demonstrated that the conformal symmetry can be applied for
the effective potential approach even at the presence of  the mass parameter.
Within the stationary phase method, it becomes possible if one introduces the special treatment of the mass terms as a kind
of interaction in an asymptotical expansion of the generating functional.

It has been shown that the multi-loop evolution equations of the effective potential, which are given by 
the corresponding anomalous dimensions,
can be derived using the results of BM-approach \cite{Braun:2013tva} together with
the original vacuum $V_{z,x}$-operation and the pole relations, 
see Eqn.~(\ref{Vac-c2}) and Eqns.~(\ref{Diam-1-2-3}), (\ref{H-3}).  
The proposed technique of computation is a new one and it involves the almost algebraic scheme
of the anomalous dimension calculations. It is also demonstrated the important role 
of ${\rm P}$-operator stemmed from the
use of pole relations for $Z$-factors, see Eqn.~(\ref{H-3}).
Indeed, the BM-method deals only with the corresponding Green functions of non-local operators
in order to extract the coefficient at $1/\varepsilon$. By definition, this coefficient determines the anomalous dimension. 
The introduced $V_{z,x}$-operation transforms the coefficient at $1/\varepsilon$ of Green functions
into the coefficient at the highest $1/\varepsilon$-singularity of the effective potential.
Then, the coefficient at the highest $1/\varepsilon$-singularity can be transformed to the 
the coefficient at the $1/\varepsilon$-singularity with the help of
the pole relations derived for the effective potential, see Eqn.~(\ref{H-3}).   

The proposed approach should be also considered as an alternative way to calculate the effective potential
within the massive $\varphi^4$-like models and it can be used for the other more complicated cases with, say,
the spontaneous symmetry breaking.

\section*{Acknowledgements}

We thank M.~Hnatic, A.~Manashov, S.V.~Mikhailov and L.~Szymanowski for very useful discussions.

%
%
%%%%%%%%%%%%%%%%%%%%%%%%%%%%%%%%%%%%%%%%%%%%%%%%%%%%%%%%%%%%%%%%%%%%%%%%%%%%%
\appendix
\renewcommand{\theequation}{\Alph{section}.\arabic{equation}}
\section*{Appendices}
%%%%%%%%%%%%%%%%%%%%%%%%%%%%%%%%%%%%%%%%%%%%%%%%%%%%%%%%%%%%%%%%%%%%%%%%%%%%%%%%%%%%%%%%%%
\section{RG-like evolution equations}
\label{App:EE}
%%%%%%%%%%%%%%%%%%%%%%%%%%%%%%%%%%%%%%%%%%%%%%%%%%%%%%%%%%%%%%%%%%%%%%%%%%%%%%%%%%%%%%%%%%

We begin this section sketching the main aspects of renormalization group (RG-like) evolution equations.
The first stage is to write the RG equation for an arbitrary (1PI) Green function $\Gamma(\{p_i\})$ in $p$-space
(the momentum space). Having used the relation between the bare and renormalized function
\begin{eqnarray}
\label{bare-ren}
&&\Gamma_0(\{p_i\}; e_0; \Lambda) = \mathbb{Z}_\Gamma(\{p_i\}; e_0(e); \Lambda, \mu)\,
\Gamma(\{p_i\}; e; \mu),
\nonumber\\
&& \text{with}\,\,\,
e\stackrel{\text{def}}{=} \big\{ g, m\big\},
\end{eqnarray}
from the trivial condition $\mu d_\mu\Gamma_0=0$, we obtain the following RG-like equation
\begin{eqnarray}
\label{rg-1}
&&
-\gamma_\Gamma \Gamma + \mu d_\mu\Gamma=0,
\end{eqnarray}
where
\begin{eqnarray}
\label{z-d}
% \mathbb{Z}_\Gamma^{-1}
&&\mu \partial_\mu \ln\mathbb{Z}_\Gamma\stackrel{\text{def}}{=}-\gamma_\Gamma,
\nonumber\\
&&
\mu d_\mu\stackrel{\text{def}}{=} \mu\frac{\partial}{\partial\mu}+\mu\frac{\partial e}{\partial\mu} \frac{\partial}{\partial e}.
\end{eqnarray}
Throughout the paper, we follow the short notations as $\partial_\mu=\partial/\partial \mu$ and so on.

In Eqn.~(\ref{bare-ren}), $\Lambda$ and $\mu$ denote the regularization parameter within some scheme and
the corresponding subtraction point, respectively. Moreover, we adhere the scheme where all renormalized parameters have been considered as
the independent set, {\it i.e.} the renormalization constants and the bare parameters are functions of the renormalized parameters
(see for details \cite{Vasilev:2004yr}).

Clearly, the formal solution of Eqn.~(\ref{rg-1}) takes the form
\begin{eqnarray}
\label{sol-rg-1}
\Gamma=C\, \text{exp}\Big\{ \int (d\ln\mu)\, \gamma_\Gamma\Big\}.
\end{eqnarray}
The next item is to make the scale-transformation for momenta, 
$p_i\to t\,p_i$, and to use the Euler theorem for homogenous function.
As a result we have the equation
\begin{eqnarray}
\label{rg-t}
\Big[
t\partial_t + e\partial_e + \mu\partial_\mu -D
\Big] \Gamma( t\{p_i\}; e, \mu)=0,
\end{eqnarray}
where $D=\text{dim}_M[\Gamma]$. Excluding the derivative over $\mu$ in Eqn.~(\ref{rg-t})
with the help of Eqn.~(\ref{rg-1}), we finally derive the evolution equation as
\begin{eqnarray}
\label{ee-f}
\Big[
t\partial_t + \big(e - \beta_e \big)\partial_ e + \overline{\gamma}_\Gamma
\Big] \Gamma( t\{p_i\}; e, \mu)=0,
\end{eqnarray}
where
\begin{eqnarray}
&&\big(e - \beta_e \big)\partial_e= -\beta_g \partial_g +
m\big(1-\gamma_m\big)\partial_m,
\nonumber\\
&&\overline{\gamma}_\Gamma = \gamma_\Gamma -D.
\end{eqnarray}
The evolution equation, see Eqn.~(\ref{ee-f}), has a solution which can be presented as
\begin{eqnarray}
\label{sol-ee-f}
\Gamma( t\{p_i\}; e, \mu) = f(t)\, \Gamma( \{p_i\}; \bar e(t,e), \mu),
\end{eqnarray}
where
\begin{eqnarray}
\label{ft}
f(t)=t^D \text{exp}\Big\{ -\int\limits^t_0 \, (d\ln\tau)\, \gamma_\Gamma(\bar g)  \Big\}.
\end{eqnarray}

The analogous RG equation (or, in another words, the evolution equation) can be written, for example, for the non-local
operator of scalar fields, see for details \cite{Braun:2013tva}, the bare operator of which reads
\begin{eqnarray}
\label{non-loc-1}
{\cal O}_0(x; z_1n, z_2n)=\varphi_0(x_1+z_1n) \varphi_0(x_2+z_2n)\Big|_{x_1=x_2},
\end{eqnarray}
where $x_i$ imply the correspondent coordinates in the position space,  $n$ is the light-cone minus basis vector,
$n\equiv(n^+,\, n^-,\, n_\perp)=(0,\, 1,\, 0_\perp)$, and $z_i$ are simple numbers.

The renormalized light-cone operator, denoted now as
(here, the renormalization constant $Z$ is in general an integral
operator acting on the coordinates $z_i$ and it has an
expansion in inverse powers of $\varepsilon$)
\begin{eqnarray}
\label{OR-def}
[{\cal O}](x; z_1n, z_2n)=Z {\cal O}_0(x; z_1n, z_2n),
\end{eqnarray}
fulfils the RG equation given by
\begin{eqnarray}
\label{rg-ee-nonloc-2}
 \Big\{ \mu\partial_\mu + \beta(g)\partial_g + \mathbb{H}\Big\} [{\cal O}](x; z_1n, z_2n) = 0,
\end{eqnarray}
where the coupling constant $g$ and
the evolution kernel (or the Hamiltonian, or the anomalous dimension) $\mathbb{H}$
can be found in \cite{Braun:2013tva}. The latter is given by
\begin{eqnarray}
\label{H-1}
\mathbb{H} =
-\Big( \mu \partial_\mu \mathbb{Z}
\Big) \mathbb{Z}^{-1} = 2g\partial_g Z_1(g) + 2\gamma_\varphi,
\end{eqnarray}
where $\mathbb{Z} = Z Z_1^{-1}$. The $Z_1$-factor is taken from the
multiplicatively renormalized standard action which is of the form 
\begin{eqnarray}
\label{Sr}
[S](\varphi)=\int (d^D x) \Big\{
\frac{1}{2} Z_1(\partial\varphi)^2 +Z_3 \mu^{2\varepsilon} g V(\varphi)
\Big\}.
\end{eqnarray}
To our goal, we study the similar RG evolution equations written for the
specially-defined Green functions, see Eqns.~(\ref{Diff-1}) and (\ref{Diff-2}).

%%%%%%%%%%%%%%%%%%%%%%%%%%%%%%%%%%%%%%%%%%%%%%%%%%%%%%%%%%%%%%%%%%%%%%%%%%%%%%%%%%%%%%%%%%
\section{On the $\delta(0)$-singularity/uncertainty }
\label{App:Delta}
%%%%%%%%%%%%%%%%%%%%%%%%%%%%%%%%%%%%%%%%%%%%%%%%%%%%%%%%%%%%%%%%%%%%%%%%%%%%%%%%%%%%%%%%%%

One of the key tools in the paper is the use of massless vacuum integrations which lead inevitably to $\delta(0)$.
Since $\delta(0)$, in the function space, represents the singularity, everyone is trying to avoid working with such an object.
However, within the sequential approach we are able to parametrize this singularity via the corresponding
meromorphic function. The point in question is that we can parametrize $\delta(0)$-singularity by many different ways.
It turns out that in our case we can fix the certain way of the $\delta(0)$-parametrization.
In this appendix we give the important details of the parametrization of $\delta(0)$-singularity within the
sequential approach to the singular generated functions (= distributions in the western literature).
%In the paper, we adhere the dimensional regularization for the renormalization.

Due to the dimensional analysis, it is not difficult to conclude that all massless vacuum integrations are identical zero, {\it i.e.}
\begin{eqnarray}
\label{V-in-1}
\mathcal{V}_n=\int \frac{(d^D k)}{[k^2]^n}=0 \quad \text{for}\,\, n\not= D/2.
\end{eqnarray}
However, the case of $n=D/2$ (or $n=2$ if $\varepsilon\to 0$) requires the special attention. Indeed,
the dimensional analysis argumentation does not now work, but the nullification of $\mathcal{V}_{D/2}$ takes still place.
It happens because the ultraviolet and infrared divergencies are cancelled each other.
Therefore, if we are interested in the ultraviolet divergencies only, $\mathcal{V}_{D/2}$ is not zero.
To demonstrate it, let us consider the corresponding integration with the IR-regularized limit in the
spherical co-ordinate system. We write the following (here, the limit $\varepsilon\to 0$ together with the Euclidian measure in $k$-space
have been assumed)
\begin{eqnarray}
\label{V-in-2}
\mathcal{V}_{2}=\int_{UV} \frac{(d^D k)}{[k^2]^2}\equiv
 \frac{\pi^{D/2}}{\Gamma(D/2)} \int_{\mu^2}^{\infty} d\beta \beta^{D/2-3} \quad \text{with}\,\,\, \beta=|k|^2,
\end{eqnarray}
where $\mu^2$ plays a role of IR-regularization and the angular integration given by the measure $d\Omega$ is calculated explicitly.
By performing the $\beta$-integration, we get that
\begin{eqnarray}
\label{V-in-3}
\mathcal{V}_{2}=
 \frac{\pi^{2-\varepsilon} \mu^{-2\varepsilon} }{\Gamma(2-\varepsilon)}  \, \frac{1}{\varepsilon} \Big|_{\varepsilon\to 0}.
\end{eqnarray}
It is a very-well known result, see for example \cite{Grozin:2005yg, Grozin:2007zz}.

On the other hand, we are able to use the vacuum integration method proposed by Gorishni and Isaev \cite{Gorishnii:1984te}
to get the different representation for $\mathcal{V}_n$. It reads
\begin{eqnarray}
\label{V-in-4}
\mathcal{V}_{n}=\int \frac{(d^D k)}{[k^2]^n} =
\frac{2i\, \pi^{1+D/2}}{(-1)^{D/2}\, \Gamma(D/2)} \delta(n-D/2).
\end{eqnarray}
Since $D=4-2\varepsilon$, we have the only contribution given by
\begin{eqnarray}
\label{V-in-4-2}
\mathcal{V}_{2}=\int \frac{(d^D k)}{[k^2]^2} =
\frac{2i\, \pi^{3-\varepsilon}}{ \Gamma(2-\varepsilon)} \delta(\varepsilon).
\end{eqnarray}
It is clear that $\delta(\varepsilon)$ is not zero. Hence, the delta-function should reflect the UV-divergency.
We specially emphasize that the representations of $\mathcal{V}_2$ given by Eqns.~(\ref{V-in-3})  and (\ref{V-in-4-2})
are equivalent.
Thus, the matching of Eqns.~(\ref{V-in-3})  and (\ref{V-in-4-2}) gives the unique way of $\delta(0)$-parametrization,
{\it i.e.}
\begin{eqnarray}
\label{D-treat}
\delta(0)=\lim_{\varepsilon\to 0}  \delta(\varepsilon)=\lim_{\varepsilon\to 0}  \frac{1}{\varepsilon}.
\end{eqnarray}

Indeed, as well-known, the delta-function is a linear singular functional (which cannot be generated by any locally-integrated functions)
defined on the suitable finite function space.
At the same time, the delta-function can be understood with the help of the fundamental
sequences of regular functionals provided the corresponding weak limit under consideration (see for example  \cite{Antosik:1973}).
Besides, one of the delta-function representations is related to the following realization
\begin{eqnarray}
\label{Delta-Real}
\delta(t)=\lim_{\varepsilon\to 0} \delta_\varepsilon(t)\equiv
\lim_{\varepsilon\to 0}  \frac{{\cal St.F.}(-\varepsilon \le t \le 0)}{\varepsilon},
\end{eqnarray}
where ${\cal St.F.}(-\varepsilon \le t \le 0)$ denotes the step-function without any uncertainties.

Returning to Eqn.~(\ref{V-in-4-2}), we observe that the treatment of $\delta(\varepsilon)$ as the linear (singular)
functional on the finite function space with $d\mu(\varepsilon)=d\varepsilon \phi(\varepsilon)$ meets some difficulties
within the dimensional regularization approach. Indeed, for the practical application,
the parameter $\varepsilon$ is not a convenient variable for the construction of the corresponding finite function space.

So, within the sequential approach \cite{Antosik:1973}, the delta-function might be considered as the usual singular (meromorphic)
function and  the $\delta(0)$-singularity/uncertainty can be treated as $\lim_{\varepsilon\to 0}  1/\varepsilon$ \cite{Anikin:2020dlh}.
For the demanding reader, the representation of Eqn.~(\ref{D-treat})  should be understood merely as a symbol.
This representation is also backed by the obvious fact that Eqns.~(\ref{V-in-3})  and (\ref{V-in-4-2})
are equivalent ones modulo the normalization which is now irrelevant.

To conclude, it is worth to notice that representation of Eqn.~(\ref{D-treat}) should be used carefully.
A very informative example can be found in \cite{Efimov:1973pjo}.

%%%%%%%%%%%%%%%%%%%%%%%%%%%%%%%%%%%%%%%%%%%%%%%%%%%%%%
%%%%%%%%%%%%%  References %%%%%%%%%%%%%%%%%%%%%%%%%%%%%%%%%%
%%%%%%%%%%%%%%%%%%%%%%%%%%%%%%%%%%%%%%%%%%%%%%%%%%%%%%

\end{document}